\documentclass[10pt,paper,twocolumn]{IEEEtran}
\usepackage[T1]{fontenc}
\usepackage{graphicx}
\usepackage{caption2}
\usepackage{array}
\usepackage{makecell,colortbl,xcolor,booktabs,multirow}
\usepackage{setspace,subfigure}
\usepackage{cite}
\usepackage{hhline}
\usepackage{diagbox}
\usepackage{url}
\ifCLASSINFOpdf
\else
\fi
\usepackage{amsmath}
\usepackage[cmintegrals]{newtxmath}
\begin{document}

\title{\LARGE Distributed Swarm Learning for Edge Internet of Things} 

\author{Yue~Wang,~\IEEEmembership{Senior Member,~IEEE}, Zhi~Tian,~\IEEEmembership{Fellow,~IEEE}, Xin~Fan,~\IEEEmembership{Member,~IEEE}, \\ Zhipeng~Cai,~\IEEEmembership{Fellow,~IEEE}, Cameron~Nowzari,~\IEEEmembership{Senior Member,~IEEE},
and Kai~Zeng,~\IEEEmembership{Member,~IEEE}
}

\maketitle
\section*{Abstract}
The rapid growth of Internet of Things (IoT) has led to the widespread deployment of smart IoT devices at wireless edge for collaborative machine learning tasks, ushering in a new era of edge learning.
With a huge number of hardware-constrained IoT devices operating in resource-limited wireless networks, edge learning encounters substantial challenges, including communication and computation bottlenecks, device and data heterogeneity, security risks, privacy leakages, non-convex optimization, and complex wireless environments.
To address these issues, this article explores a novel framework known as distributed swarm learning (DSL), which combines artificial intelligence and biological swarm intelligence in a holistic manner.
By harnessing advanced signal processing and communications, DSL provides efficient solutions and robust tools for large-scale IoT at the edge of wireless networks.

\IEEEpeerreviewmaketitle

\section*{Introduction}
\noindent
Smart Internet of Things (IoT) become the workhorse at wireless edge, where IoT devices gather valuable data directly from edge environments and fuel the recent trend of artificial intelligence (AI) applied at the edge, a.k.a. edge learning.
However, traditional machine learning (ML) methods are not suitable for edge learning, since they hinge on collecting raw data  from local devices and raising concerns of privacy leakages and security risks. Alternatively, federated learning (FL) has emerged to allow distributed learning while keeping data locally~\cite{mcmahan2017communication}, which has led to fruitful attempts in implementing AI among  distributed terminal users, e.g., cell phones.

The success of vanilla FL typically relies on ideal learning conditions and perfect wireless environments, where cell phones possess powerful computing capability 
and communicate over secure networks with stable connectivity\cite{mcmahan2017communication}.
But, these assumptions become invalid in many real-world edge IoT applications, where low-cost IoT devices typically have constrained computation and communication capability.
Further, when neural networks are employed, the large size of model parameters poses a major challenge in transmitting all the updates from distributed workers during the training stage of FL.
In addition, stochastic gradient descent (SGD) is widely used for model training, assuming independent and identically distributed (i.i.d.) data samples at local workers. 
This assumption however does not hold in edge IoT, as IoT devices only access small-volume data.
All these factors give rise to the following challenges to IoT-driven edge learning.

{\color{black}
{\em Challenge-1: Communication bottleneck}. The large size of learning models and huge number of distributed workers incur high communication costs for edge learning in IoT systems, while low-cost IoT devices are prone to power constraints, bandwidth limits, and communication resource scarcity.

{\em Challenge-2: Non-convex optimization}.
Gradient-based algorithms get trapped in local optima when tackling non-convex problems, e.g.,
training 
neural networks with nonlinear activation. 
This problem worsens in distributed learning, particularly in IoT scenarios where edge devices access limited data.

{\em Challenge-3: Data and device heterogeneity}.
Edge learning faces statistical heterogeneity in local training data across workers, also known as the non-i.i.d. data issue, as well as device heterogeneity in IoT hardware capability and link quality, which degrades edge learning performance significantly.

{\em Challenge-4: Privacy and security concerns}.
Although standard FL excels in idealized networks 
and attack-free learning scenarios, it proves susceptible to outliers, eavesdroppers, privacy breach, and Byzantine attacks, all of which inevitably exist in IoT applications, wireless systems, and edge networks.

{\em Challenge-5: Complex network environments}. Edge IoT systems usually experience communication errors, node failures and link interruptions. Further, 
for large-scale networks, workers have to deal with time-varying topology and environments where traditional designs for the static case fail to work.}

While some of the aforementioned challenges have been partially addressed in recent literature on FL for IoT~\cite{khan2021federated}, the primary focus and main efforts are found on revising, adapting, and customizing standard FL methods.
Most existing results stem from the AI perspective alone, but neglect unique characteristics of edge IoT, including the large population of IoT devices, limited capability of each IoT device, and  local 
data of small volume.
Intriguingly, biological organisms in nature having swarm intelligence, even if individually weak, have successfully demonstrated superior strength in collectively searching the optimal solutions, recovering from errors, and adapting to environment changes. Since these attributes are sought out by IoT-driven edge learning, biological intelligence (BI) is expected to boost efficiency and robustness of edge learning among massive low-cost IoT.

{\color{black}
This article investigates a holistic integration of AI and BI, by leveraging swarm intelligence and cooperative gains among massive IoT workers to develop efficient and robust distributed learning techniques at wireless edge. A new edge learning paradigm, called distributed swarm learning (DSL)~\cite{JSTSP2022DSL}, is proposed by connecting the AI-enabled FL with the BI-inspired particle swarm optimization (PSO)~\cite{kennedy1995particle}, as shown in Fig.~\ref{fig:Fig_1}.}
{\color{black}While the discussions in this article mainly refer to the vanilla FL for clarity, the DSL framework allows the PSO-related techniques to be combined with the general FL family,
to address the key challenges and unique characteristics of edge IoT systems with a large number of resource-constrained devices.
\begin{figure}
  	\centering 		
  \includegraphics[width=3.5in]{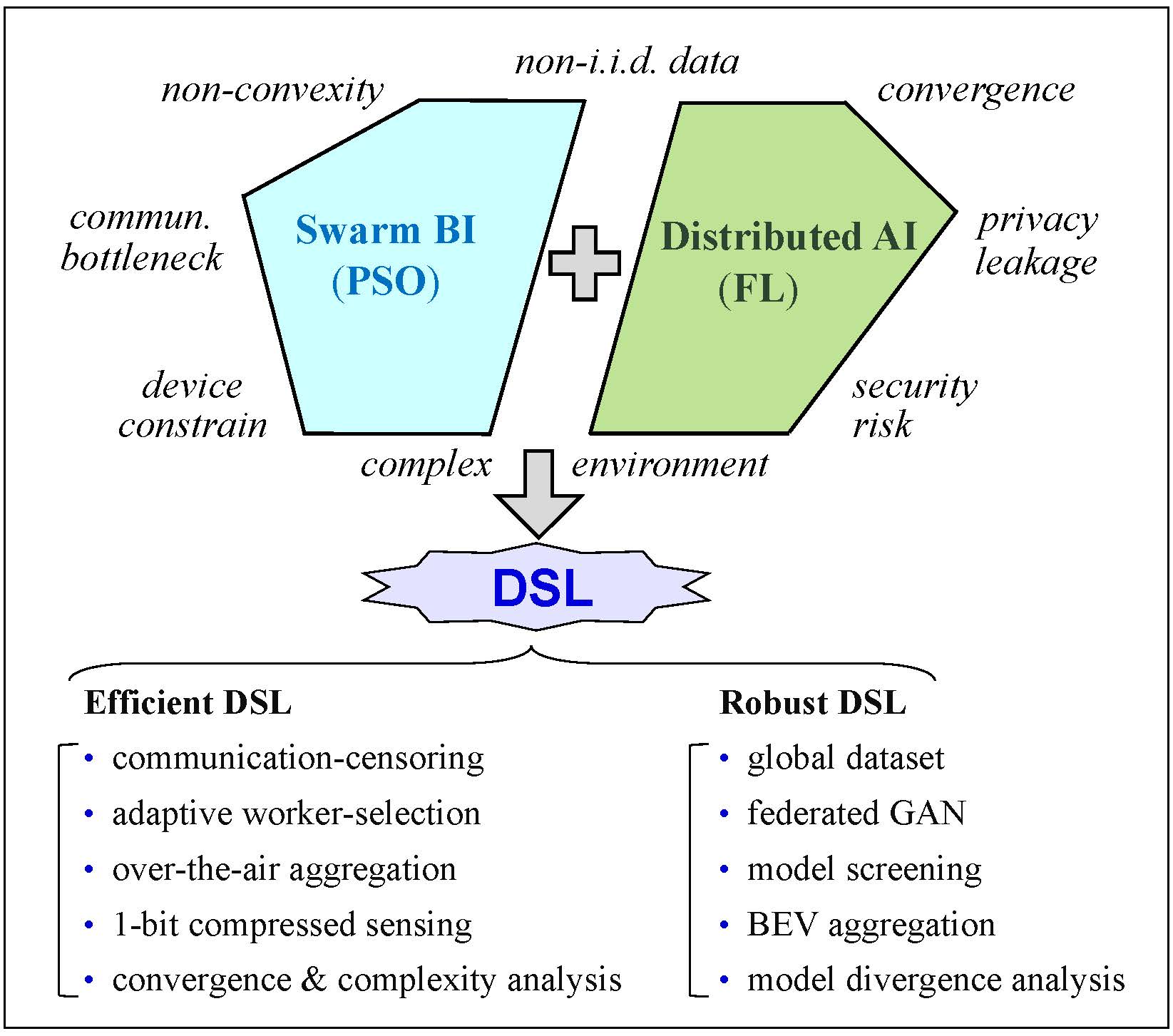}
  \caption{\small {\color{black}Distributed swarm learning framework for edge IoT.}}
  \label{fig:Fig_1}
\end{figure}

First, to overcome {\em Challenge-1}, DSL judiciously selects only a small number of  workers to share their local model updates for collaboration in the swarm. Thus, the communication costs and transmit power consumed in DSL can be saved dramatically. Further, communication efficiency of DSL can be strengthened in combination 
with parsimonious transmission methods and strategies. 

Second,
DSL combines  the velocity in PSO with the gradient in SGD, which amounts to embedding a bio-inspired mechanism to FL. 
Thanks to the exploration-exploitation mechanism of PSO in velocity updating, particle solutions in swarm have an increased chance
to escape from the local optimum traps in {\em Challenge-2},
given heterogeneous data and devices in {\em Challenge-3}. Thus, iterative DSL algorithms converge faster to better accuracy at reduced overall communication/computation rounds than PSO or FL alone.

Third, by using over-the-air aggregation, multiple workers are allowed to simultaneously transmit their local model variables over the same time-frequency resources. Hence, device restrictions in communication bandwidth and resource constraints in {\em Challenge-1} would minimally affect the DSL operations. Further, over-the-air aggregation enhances  privacy preserving capability due to the inherent inaccessibility to individual local updates, which mitigates the risk of potential model inversion attacks in {\em Challenge-4}.

Fourth, introducing a very small-volume global dataset plays a key role in providing high accuracy, efficiency and robustness of DSL in the presence of non-i.i.d. data in {\em Challenge-3} and adversarial attacks in {\em Challenge-4}. A part of this globally shared dataset is used for training, whose effectiveness in alleviating 
the non-i.i.d. problem is proved through model divergence analysis. The other part 
is used to calculate the fair-value loss for scoring the 
models, which not only helps 
in worker selection but also enables 
to screen potential 
attacks.

Last, in response to {\em Challenge-5}, per the efficient and robust updating rule of DSL established in swarm, even if some workers do not respond to the request for sharing local updates, the rest workers are still able to work as long as the network connectivity is not broken. This enhances the robustness of DSL against node and link failures happened in edge IoT scenarios. DSL has potentials to be implemented in massively parallel and adaptive manners, and so scale better than traditional distributed approaches.
}


\section*{Framework of Distributed Swarm Learning}
\noindent
{\color{black}This article refers to
IoT-driven edge learning paradigms, where massive low-cost IoT workers collaboratively learn 
common model parameters form distributed local data,  
given limited communication, computation, and storage capabilities 
at each worker. Further, the learning performance is impacted by bandwidth limits, malicious attacks, node malfunctions, link failures,
heterogeneity in devices and data, and complex network environments.
}

\subsection*{Federated Learning}\label{sec:FL}
\noindent
Standard FL is originally designed in ideal learning and network settings \cite{mcmahan2017communication}, without the additional constraints on IoT devices and edge environments.
Therein the minimization of the loss function is typically carried out by the SGD algorithm, whose performance highly relies on i.i.d. data samples and error-free transmissions. At workers, they individually update their local models, and then the global model or averaged gradient is calculated from all local updates, which works as the initial point for local updating at the next iteration. Computation and communication take place at all local workers in every iteration, until FL convergence.
In the case of non-convex problems, existing gradient-based FL solutions may converge to undesired local optima, but without effective mechanisms to escape from such traps.

\subsection*{Particle Swarm Optimization}\label{sec:PSO}
\noindent
Inspired by the BI behavior of swarming animals, PSO is capable of solving complex optimization problems without requiring convexity assumptions~\cite{kennedy1995particle}.
In PSO, mimicking swarm behavior in animal flocks, particles run the movements of a swarm to collectively search for an optimal solution. Each particle's position represents a potential solution to the problem, while its velocity dictates the direction for updating in the following step.
To discover the globally optimal solution within the swarm, particles collaboratively update their velocities and positions over iterations. Notably, the velocity update is achieved through a weighted combination of three components: inertia from the previous direction, individual guidance toward each particle's historical best variable, and social guidance toward the globally best variable discovered by the entire swarm. Such a weighted combination introduces an effective mechanism for balancing self-exploration and group-exploitation in swarm optimization. 

\subsection*{Distributed Swarm Learning}\label{sec:DSL}
\noindent
For model updating, the gradient in FL 
and the velocity in PSO 
play a similar role as the updating direction, but their updating principles are different. In PSO, the velocity is a weighted combination of three sub-directions, where the collaboration is reflected in the socially updated sub-direction towards the up-to-date globally best variable. As a gradient-free stochastic optimization method, PSO excels in its ability to collaboratively search for the globally optimal solutions to complex problems, thanks to its built-in exploration-exploitation mechanism and the swarm nature. 
But, PSO assumes a globally common loss function for all workers to facilitate collaboration, which no longer holds in distributed learning problems characterized by data-dependent local loss functions. In contrast, FL is a gradient-based learning algorithm with fast convergence. However, it is subject to local optimum traps and suffers from high communication costs in massive IoT systems. 

The advantages and disadvantages mentioned for FL and PSO motivate to establish a connection between distributed learning and swarm optimization techniques. This integration aims to harness the strengths of both artificial and biological intelligence for enhanced problem-solving capability.
There exist few recent efforts to directly employ PSO concepts to enhance the performance of FL.
In~\cite{qolomany2020particle}, FL is applied in the learning process, whereas PSO is used separately for the purpose of searching the optimal hyperparameters. In~\cite{park2021fedpso}, PSO and FL are combined in a straightforward manner for an idealized distributed setting with i.i.d. data among workers and in the absence of any security attacks. 
The work~\cite{park2021fedpso} builds on an implicit assumption that a common loss function is accessible to all local workers, simplifying the process of evaluating the globally best model through a simplistic single-worker selection scheme. However, the loss function is only partially observable at local workers, which is data-dependent and therefore varies across workers in distributed learning cases. Thus, the method of~\cite{park2021fedpso} may not perform effectively in edge IoT scenarios where non-i.i.d. data is encountered at distributed local workers.

To fill the identified technical gaps above,
DSL is recently proposed as
a novel distributed learning framework tailored to edge IoT~\cite{JSTSP2022DSL},
which is schematically illustrated through the following model updating step per iteration:
{\color{black}
\begin{align}\label{eq:DSL_w}
 &\mathbf{w}_{i,t+1}\!=\!\mathbf{w}_{i,t}+ \nonumber\\
 &\;\underbrace{\lambda_t(c_0 \mathbf{v}_{i,t}\!+\!c_1(\mathbf{w}_{i,t}^p\!-\!\mathbf{w}_{i,t}\!)\!+\!c_2 (\mathbf{w}_t^g\!-\!\mathbf{w}_{i,t}\!))}_{\mathbf{BI}}\underbrace{-{(}1{-}\lambda_t\!{)}\alpha\nabla F_i(\mathbf{w}_{i,t};\mathcal{D}_i\!)}_{\mathbf{AI}}\, ,
\end{align}
where $\lambda_t \in [0,1]$ plays an adaptive tradeoff between the BI contributions to escape from local optimum traps via swarm intelligence and the AI contributions to learn the underlying mapping function from local training data $\mathcal{D}_i$. Intuitively, a larger $\lambda_t$ encourages workers to update their models by utilizing the exploration-exploitation gains from the swarm, whereas a smaller $\lambda_t$ suggests to update models by mainly following the guidelines from the gradients with learning rate $\alpha$. Further, among the three PSO coefficients, $c_0$ is a positive number while $c_1$ and $c_2$ are positive and random (say, uniformly distributed for stochastic optimization). Such three coefficients jointly determine a combination of three sub-directions of the PSO terms, where $c_0$ can be set linearly decreasing over iterations to tune the solution search process from exploration to exploitation, $c_1$ and $c_2$ indicate the random exploration level at individual workers and the exploitation level in swarm, respectively. Thus, $c_0$, $c_1$, and $c_2$  balance exploration of individuals and exploitation in the swarm.
}

In this sense, for each worker $i$, its local model parameter vector $\mathbf{w}_{i,t}$ is updated through local velocity $\mathbf{v}_{i,t}$ and gradient $\nabla F_i$ terms jointly in \eqref{eq:DSL_w}, where $\mathbf{w}_{i,t}^p$ records its own historical best variable, $\mathbf{w}_{t}^g$ is the globally best variable of the swarm,  $F_i(\cdot)$ is the local loss evaluated on the local dataset $\mathcal{D}_i$. 
{\color{black}Upon updating $\mathbf{w}_{i,t}$, at each worker, the locally best variable $\mathbf{w}_{i,t}^p$ and the fair-value loss $F_{i,t}^p$ 
are calculated by using a small global dataset $\mathcal{D}_g$ that is introduced and made available to all workers: 
\begin{equation}\label{eq:DSL_wp}
      \left\{\mathbf{w}^p_{i,t}\, ,  F_{i,t}^p\right\}  =  \min\left\{ F_{i,t-1}^p\, , F_i(\mathbf{w}_{i,t}; \mathcal{D}_g) \right\}.
    \end{equation}
To do this, each edge device only needs to locally store the previous best model $\mathbf{w}_{i,t-1}^p$ and its associated loss function value $F_{i,t-1}^p$. 
These two quantities about the historical information are locally maintained by either overwriting the memory as $\mathbf{w}^p_{i,t} = \mathbf{w}_{i,t}$ and $F_{i,t}^p = F_i(\mathbf{w}_{i,t}; \mathcal{D}_g)$ when a better local solution is reached with a lower loss i.e., $F_i(\mathbf{w}_{i,t}; \mathcal{D}_g) < F_{i,t-1}^p$, or simply keeping the previous ones i.e., $\mathbf{w}^p_{i,t} = \mathbf{w}^p_{i,t-1}$ and $F_{i,t}^p = F_{i,t-1}^p$ otherwise. As a result, such memory overhead is quite small for maintaining the historical models and function values at each local device.
}

When a central server is present,
it collects these scalar loss values $F_{i,t}^p \, , i=1, \dots, U$ 
from all $U$ workers to perform adaptive worker selection, so that only a few (say $S_t$) workers with the lowest losses are selected to transmit their locally best variables $\mathbf{w}_{i,t}^p$ to the server for updating the global $\mathbf{w}_{t}^g$ via over-the-air 
aggregation: 
{\color{black}
\begin{equation}\label{eq:DSL_wg}
\mathbf{w}_{t}^g=\frac{1}{S_t}\sum_{i\in \mathcal{S}_t} \underbrace{\mathbf{p}_{i,t}\odot \mathbf{h}_{i,t}\odot}_{\textbf{Over-the-air}} \mathbf{w}^p_{i,t}+\mathbf{n}_{t}\, ,
\end{equation}
where $\odot$ is the Hadamard product,  $\mathbf{n}_{t}$ is additive noise, $\mathcal{S}_t$ of size $S_t=|\mathcal{S}_t| \, (1 \leq S \leq U)$ is a subset of selected workers, $\mathbf{p}_{i,t}$ and $\mathbf{h}_{i,t}$ represent the transmit power and channel state at the $i$-th worker,  respectively.
When channel states are known, power control and worker selection offer the freedom for design tradeoffs, which can be carried out via joint optimization for worker selection and resource allocation~\cite{fan2021joint}.
Meanwhile, over-the-air analog aggregation not only improves communication efficiency, but also enhances the data privacy preserving capability thanks to the inherent
inaccessibility to individual local updates.
}
\section*{Efficient Distributed Swarm Learning}
\noindent
A body of literature has been devoted to address the communication efficiency of FL, via gradient sparsification, message quantization, and infrequent transmissions of local model updates. They either compress the information to be transmitted or drop less-informative transmissions prior to aggregation of local updates.
On the other hand, all of these methods still entail all participating workers to exchange their local model updates, which is however not well-suited for edge IoT systems and leads to tremendous communication costs, especially when dealing with a large number of IoT devices. To save device energy consumption and reduce total communication cost in edge networks, DSL incorporates parsimonious transmission and aggregation schemes with the built-in exploration-exploitation mechanism in a swarm.

\subsection*{Efficient Communication and Computation}
\noindent
Recall the different collaboration mechanisms between FL and PSO: In FL, all workers report their local updates for global averaging; while in PSO, only one worker is selected as the up-to-date global best to share its local update with others. For the purpose of the best worker selection in PSO, each worker is also requested to share the minimum value of its historical local loss functions to compare with that of others.
FL and PSO work as the two extreme cases under the DSL framework. To collect the benefits from both sides, the communication and aggregation protocols in DSL need to be carefully designed with theoretical backing.

\subsubsection*{Communication Censoring}
In DSL, all the workers report their individual historical best local function values, based on which a few best workers with the lowest losses are selected for the calculation of the global model variables in each iteration as \eqref{eq:DSL_wg}.
Even though the loss function values are scalar, such a reporting process still consumes large communication overhead as the total number of workers goes large in edge IoT. Note that the reported values are sorted for worker selection, while the sorting operation accommodates some level of value approximations. It prompts to introduce a communication-censoring strategy at each worker to save individual communication cost. That is, each worker assesses how much its current local best function value differs from its previously recorded one, and reports only when the difference is large enough to exceed a censoring threshold. Otherwise, the worker does not report to save transmission, and worker selection is made based on its previously reported value as an approximation of its current value. The censoring process is autonomous based on the judiciously designed local censoring threshold to guarantee convergence~\cite{xu2021coke}.

\subsubsection*{Adaptive Multi-Worker Selection}
After collecting or approximating workers' historical best local function values through the communication censoring scheme, it is essential to deploy a multi-worker selection strategy by deciding $S_t$, the number of selected workers who will contribute their local historical best variables to the global variable updating in \eqref{eq:DSL_wg}. Note that DSL encourages local workers to move along the social direction towards the global best variable as a means of exploitation of swarm collaboration in \eqref{eq:DSL_w}. To effectively fulfill the exploration-exploitation mechanism with communication efficiency, an adaptive strategy can be employed to increase the value of $S_t$ along the iterations. Intuitively, a smaller $S_t$ at the early stage of iterations allows individual workers to focus on exploring the solutions locally at little communication overhead, and then a larger $S_t$ during later iterations brings more workers to contribute to the collective swarm intelligence for high model training accuracy at convergence.

{\color{black}
On one hand, when worker selection is involved, DSL does require double iterations of communication for each round (one to communicate the loss values for worker selection and the other to communicate the local variables of selected workers for global model updating). Interestingly, such a strategy is designed to reduce the overall communication overhead and time complexity, which turns out to a reduction of the total energy consumption.
Note that the loss value is scalar, which consumes little cost to communicate compared with communicating the model variables. This extra worker selection step allows to pick a small number of most effective devices to communicate, which drastically reduces the total cost of communicating the models of all devices in the absence of this extra step.  Alternatively, if worker selection is made randomly or naively as in vanilla FL, these selected devices would not optimally contribute to model updating; in turn, it may take longer time
to converge and hence incur higher communication cost until convergence.
Overall, DSL employs a low-cost worker selection step to alleviate the heavy cost of the model communication step, thus achieving overall communication and energy efficiency.
}

\subsubsection*{Over-The-Air Analog Aggregation}
Recently, a promising technique emerges as FL over-the-air \cite{fan2021joint}. This approach leverages the fact that the model-aggregation operation in FL aligns with the waveform-superposition property of wireless analog multi-access channels (MAC).
In DSL, the global variable updating in \eqref{eq:DSL_wg}
only requires the averaged updates of $S_t$ selected workers rather than their individual local variables.
{DSL over-the-air} works in a direct and efficient way of implementing global variable updating.
It employs
analog aggregation based transmission and enables the $S_t$ selected workers to simultaneously transmit their local variables over the same time-frequency resources. This is achieved when the aggregated waveform effectively represents the averaged updates, through proper transmit power control\cite{fan2021joint}. Thus, the total consumed bandwidth is minimized and independent of $S_t$. Since $S_t$ reflects the degrees of freedom in DSL to control the exploration-exploitation mechanism in swarm intelligence, adaptive multi-worker selection strategy can be co-designed and assessed based on the impact of the over-the-air transmission on the learning performance. In addition, DSL can further incorporate other efficiency-enhancing strategies such as 1-bit compressed sensing,
which amounts to combining compression, quantization and concurrent transmission to attain impressive efficiency \cite{fan20211}.

\subsection*{Convergence Behavior}
\noindent
DSL offers superiority in solving non-convex  problems, which mainly benefits from the inherent capability in escaping from local optimum traps owing to the random weights and exploration-exploitation mechanisms embedded in PSO. Such a benefit of swarm intelligence are intuitive and widely appreciated. From the theoretical side, the global convergence properties of PSO have been of interest to the computational science community~\cite{Clerc2002particle}. While the superiority of PSO lies in its high probability of achieving global convergence, PSO itself does not necessarily improve the order of convergence speed, unless joint optimization for parameter selection and resource allocation is carried out at the system level. In convergence analysis of DSL, the metric of interest is the expected convergence rate of local workers, which is used to evaluate the convergence of stochastic algorithms. A closed-form expression for the expected convergence rate achieved by DSL is derived as a function of the number of communication rounds $T$ and the PSO-related exploration-exploitation parameters in~\cite{JSTSP2022DSL}, which is bounded on the order of $\mathcal{O}({1}/{T})$.

\subsection*{Joint Optimization of Communication and Computation}
\noindent
The convergence behavior of DSL unveils a fundamental connection between edge communications and distributed learning. It offers a fresh perspective to measure how the parameter choice of wireless systems and the hyperparameter design of computational algorithms affect the performance of edge learning. Guided by the theoretical results from convergence analysis, joint optimization of learning parameter determination, worker selection, and transmit power control is formulated as minimizing the loss function subject to the limited transmit power and bandwidth.
It amounts to a network resource optimization problem that minimizes the learning error subject to the maximum power constraints of low-cost IoT devices~\cite{fan2021joint, fan20211},
which yields the jointly optimized worker selection and power control variables for DSL operations.

{\color{black}
\subsection*{Computational Complexity}
\noindent
At the first glance, DSL does incur extra computation than vanilla FL, when DSL integrates
the velocity in PSO with the gradient in SGD
in~\eqref{eq:DSL_w}. But, the added computation and storage overheads are indeed quite low. 
This is because FL is a first-order algorithm involving relatively expensive gradient computation, while PSO is a zeroth-order algorithm that computes functional values only.
Further, the update of the locally best variable and the corresponding fair-value loss can be implemented by simply comparing the function value of the current model variable with that of the previous best loss function value in~\eqref{eq:DSL_wp} whose computational cost is near negligible.
In addition, the global model update of DSL in~\eqref{eq:DSL_wg} is implemented in a way similar as FL via analog aggregation but only over selected workers,
which does not incur extra computation at the minimum requirement on communication bandwidth.
Thus, the computational cost of DSL is dominated by gradient-related term, with little overhead from PSO-related terms.
Finally, the theoretical results on convergence analysis indicate that DSL can expedite the convergence time than vanilla FL~\cite{JSTSP2022DSL}. As a result, the overall number of computation rounds is reduced, which leads to a reduction in the computational complexity of DSL, compared to vanilla FL.
}

\section*{Robust Distributed Swarm Learning}
\noindent
In edge IoT networks, the local data samples of small volume typically turn out to be non-i.i.d. across workers. Further, there exist malfunctioning workers and even malicious attackers. In addition, link failure contributes to unreliable transmissions. All these issues call for robust measures for DSL.

\subsection*{Global Dataset Generation}
\noindent
As a vital component of the DSL framework, a globally shared dataset is introduced  in \eqref{eq:DSL_wp}
to play dual roles: to compute fair-value scores of local models for multi-worker selection and to alleviate the non-i.i.d. issue. Centralized data sharing methods would collect raw data samples from individual workers to form global datasets \cite{tuor2021overcoming}, but raise privacy concerns.
To keep raw data private at local workers, robust data augmentation methods are developed to form a global dataset, based on federated generative adversarial network (FedGAN) by training a neural generator in a distributed manner~\cite{Li2022IFL}.
This generator can be used to generate synthetic data samples as the globally shared dataset for DSL. {\color{black}Experiments show that even a very small amount of global data (say $1\%$ of the entire datasets) is adequate to deliver the expected high learning performance for robust DSL~\cite{JSTSP2022DSL}.
 This observation is consistent with other works in the literature employing a global dataset to improve robustness in FL.
The global data can be shared prior to DSL operations, and the resources needed for sharing and storage are low due to the small volume of such data.}

\subsection*{Robust Measures against Non-i.i.d. Issues}
\noindent
Having generated a global dataset via the FedGAN-based data augmentation, the global dataset is further divided into two parts: one for model training to mitigate the  non-i.i.d. challenge, and the other for scoring worker quality against Byzantine attacks, respectively. To alleviate the non-i.i.d. issue, a globally shared training dataset is distributed among all local workers. This simple idea turns out to be effective, as evaluated by a statistical analysis approach that is developed to evaluate the impact of model weight divergence on the learning performance of DSL~\cite{JSTSP2022DSL}.
The model weight divergence is measured in terms of the distance from the non-i.i.d. data distributions of local workers to that of the overall data population. The non-i.i.d. data problem can be further tackled by employing a regularized total variation in the local loss function.
Such a regularized penalty term enforces local variables to be close to each other by reducing the distance from current local updates
to the previous global variables,
which alleviates the impact from non-i.i.d. datasets.

\subsection*{Robust Measures against Byzantine Attacks}
\noindent
{\color{black}The second part of global dataset is used to score the local loss values in \eqref{eq:DSL_wp}. 
Meanwhile, such scoring dataset can also serves to screen out 
potential Byzantine attacks where attackers 
cheat in local scoring and report mismatched local model updates.
DSL will throw away the contaminated model, when the deviation of its loss value based on the scoring dataset from the averaged score of the selected workers in their reporting stage larger than 
a predefined tolerance threshold.
Then, DSL keeps inviting other workers from the swarm by following certain reselection strategy
    until the deviation is acceptable.}
Analog aggregation transmission is known to enhance the privacy preserving due to the inherent inaccessibility to the individual local model updates.
While DSL over-the-air closes the doors to model inversion attacks, it still exposes vulnerabilities to adversaries who can perform Byzantine attacks.
The work \cite{fan2022bev} indicates that even a single instance of Byzantine fault has the potential to disrupt distributed learning via analog aggregation. The impacts of Byzantine attacks to DSL are studied under the swarm setting at the edge, taking into account of the integrated nature of AI and BI in DSL~\cite{JSTSP2022DSL}. Based on the theoretical analysis~\cite{fan2022bev}, a best effort voting (BEV) based transmission power control policy can be applied to against Byzantine attacks for DSL over-the-air. 

\subsection*{Robust Measures against Node/Link Failures}
\noindent
Thanks to the analog aggregation transmission as \eqref{eq:DSL_wg}, the worker selection is not subject to constrained communication bandwidth, which allows the selection of more workers to collectively update the global model variables at no extra bandwidth. Leveraging such multi-worker diversity, DSL can enhance the resiliency to unreliable transmission links and malfunctioning workers. Further, to cope with wireless fading, truncated channel inversion method is developed for FL over-the-air~\cite{zhu2019broadband}. When some workers experience deep fading channels, the truncated channel inversion method is employed to weed out those workers with low channel gains. For DSL over-the-air, the truncated channel inversion policy needs to be designed with attention to the incorporation of the BI terms. By jointly determining the truncation thresholds with the multi-worker selection scheme in DSL over-the-air, the truncation rules can effectively balance both the communication and computation aspects of distributed learning over wireless fading channels.

\section*{Performance Evaluation}
\begin{figure}
\centering
{\includegraphics[width=2.6in]{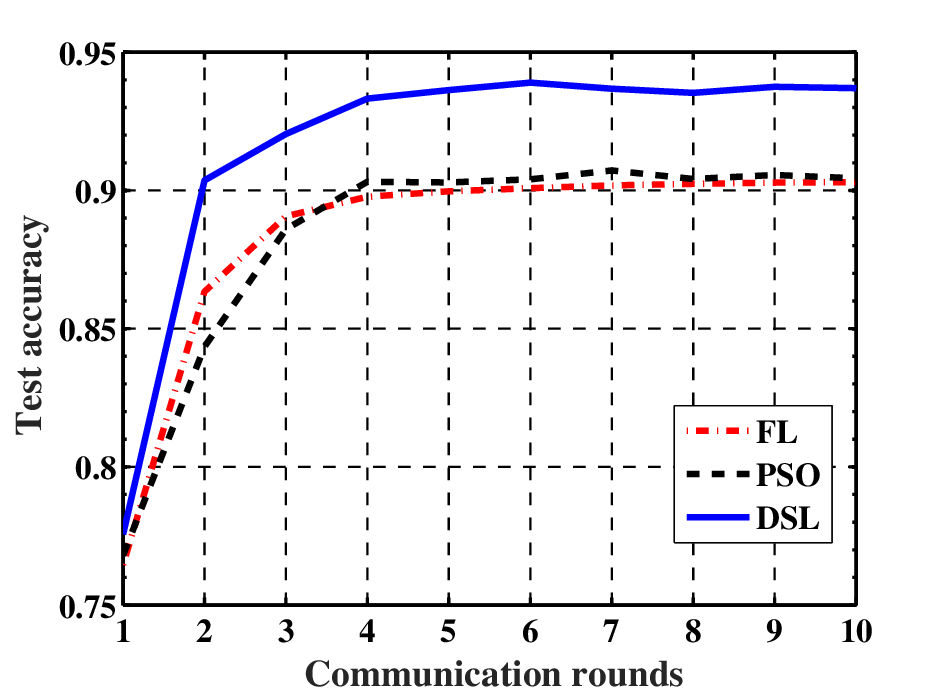}}
 \caption{\small Performance in i.i.d. cases.}
 \label{fig:iid}
\end{figure}

\begin{figure}
\centering
{\includegraphics[width=2.6in]{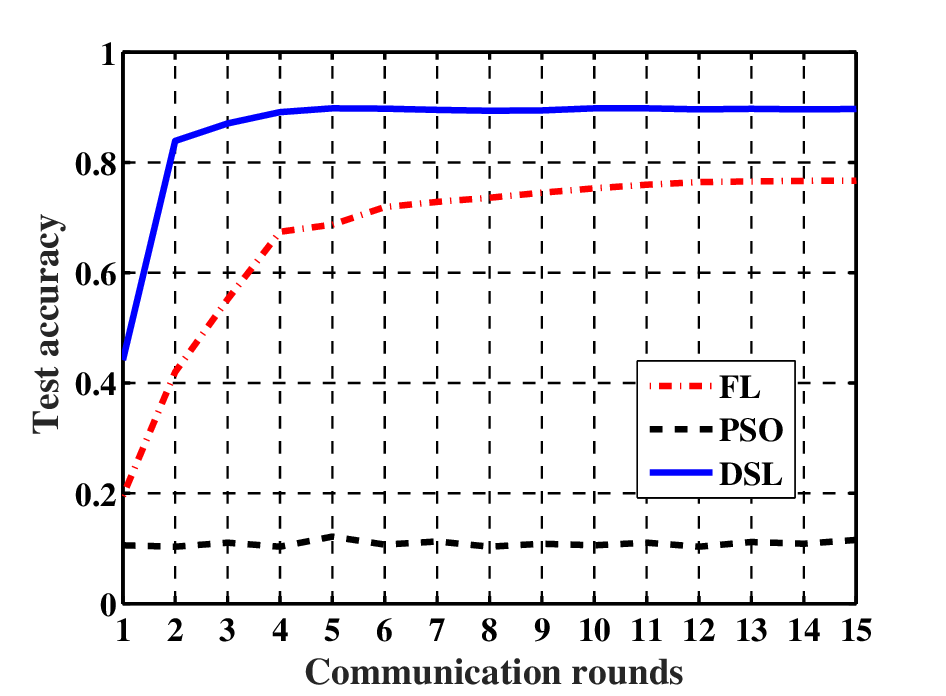}}
 \caption{\small Performance in non-i.i.d. cases.}
 \label{fig:noniid}
\end{figure}

\begin{figure}
\centering
{\includegraphics[width=2.6in]{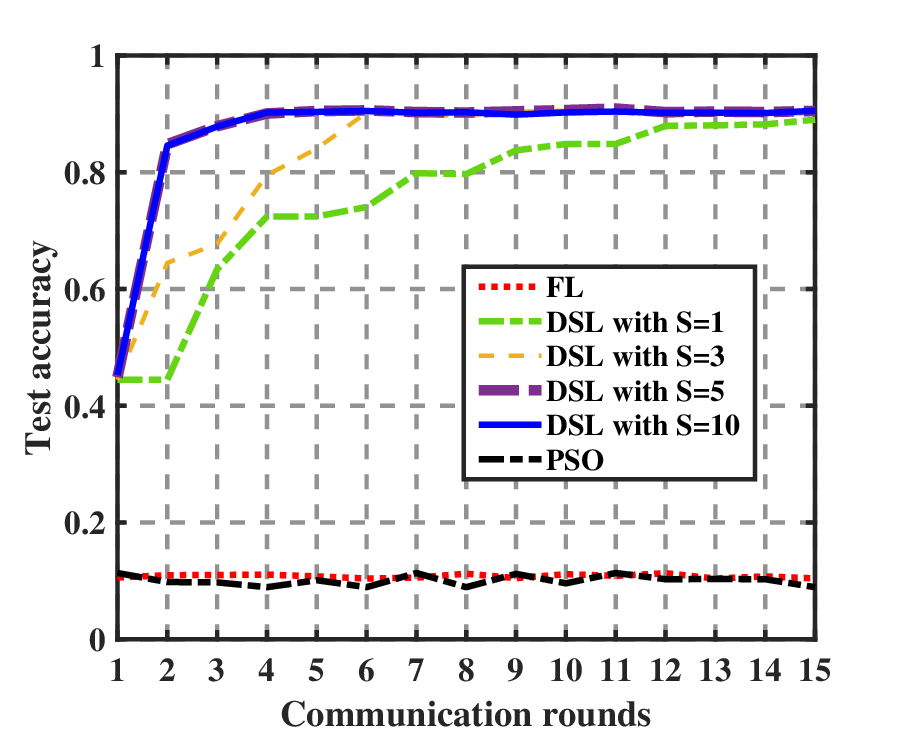}}
 \caption{\small Performance under
 Byzantine attacks and node/link failures.}
 \label{fig:attack}
\end{figure}

\noindent
For performance evaluation, DSL has been tested with $50$ distributed workers on a handwritten-digit classification task using the widely-used MNIST dataset~\cite{JSTSP2022DSL}, compared with vanilla FL and PSO as the benchmark methods.
The common learning model is a five-layer Convolutional Neural Network with $44426$ trainable parameters, using a cross-entropy loss function. As shown in Figure~\ref{fig:iid}, DSL outperforms FL and PSO in achieving the highest accuracy.
At the same performance level, the communication round of DSL is less than that of FL or PSO.
Such advantage of DSL becomes more obvious in the non-i.i.d. case in Figure~\ref{fig:noniid}, where PSO fails to work properly because its underlying assumption of a common loss function for all workers can only be approximately true in the i.i.d. case.
Further, DSL can not only effectively defend Byzantine attacks by proactively screening and removing potential attackers, but also work robustly in the presence of node/link failures by taking advantage of the diversity gain via multi-worker selection, as shown in Figure~\ref{fig:attack}.
These simulation results verify that the integration of AI and BI leads to evident improvement on learning accuracy, communication efficiency, and system robustness. This brings significant benefits to distributed learning for edge IoT, particularly when dealing with swarm devices having limited communication/computation capability and given their small-volume and heterogeneous local data at the edge of wireless networks.


\section*{Conclusion and Future Work}
\noindent
This article studies a new DSL paradigm for edge IoT systems, by bridging federated learning with particle swarm optimization. With theoretical backing, efficient information extraction and exchanging schemes are designed for high efficiency in communication and computation of model updates. Robust DSL solutions are developed to cope with data heterogeneity, node/link failure and Byzantine attack. 
While DSL provides promising improvement over existing methods, some open issues and future directions are summarized below.
\subsubsection*{Decentralized Topology}
For DSL adopting a star topology, the central server plays a pivotal role 
in coordinating the iterative model updating algorithms 
at distributed workers. 
But, it is more challenging for fully decentralized networks, 
because every worker needs to make autonomous decisions regarding the global learning task, via communication with adjacent workers only~\cite{Xu2023QC}.
There is a growing interest in decentralized DSL to capitalize on its inherent resilience to node/link failures and its ability to handle asynchronous computing in heterogeneous environments.
\subsubsection*{Security Provisioning}
Research on DSL has just started to seek understanding of system-level design and algorithm-level development, with little effort on security provisioning yet. DSL over-the-air  leaves edge learning systems still vulnerable to malicious attacks, because there is no mechanism to retrieve individual messages that help to identify attackers. 
For proactive security provisioning, it is urgent to utilize cybersecurity techniques in DSL, such as authentication and cryptographic mechanisms, while keeping the overhead and complexity low.
\subsubsection*{Efficiency-Robustness Tradeoff}
Due to the natural tradeoff between efficiency and robustness, the adopted robust aggregation measures may affect the convergence behavior of iterative DSL algorithms. For DSL over-the-air, multiple selected workers do not consume extra bandwidth, but do increase the total  transmit energy in the DSL systems. 
Understanding of such tradeoff is vital, which sheds light on the impacts of AI learning and BI optimization parameters on DSL convergence behavior.
\subsubsection*{Prototype Development}
\begin{figure}
\centering
{\includegraphics[width=2.5in]{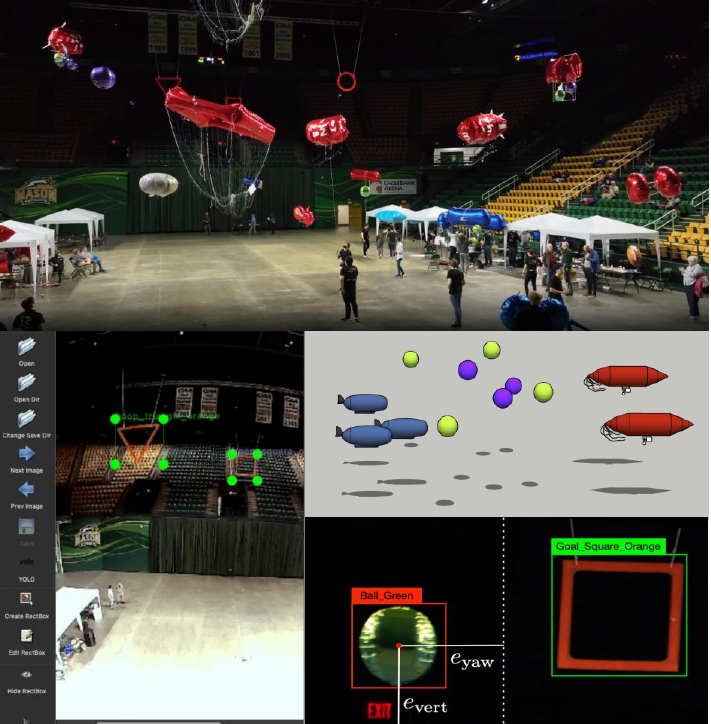}}
 \caption{\small {\color{black}LTA robots for national DTR robotic blimps competitions.}}
 \label{fig:SPARX}
\end{figure}

{\color{black}DSL prototype aims at implementation on the swarming lighter-than-air (LTA) robots designed for the national Defend The Republic (DTR) robotic blimps competitions. As shown in Figure~\ref{fig:SPARX}, in an aerial soccer game, two teams play against each other in a head-to-head match, where fleets of autonomous robots must detect and then capture neutrally buoyant balls and finally move them through goals suspended from the ceiling.
%
%
Given the onboard hardware limitation for massive LTA robots,
DSL is a good candidate solution 
to collaboratively train learning models for detecting and classifying objects, e.g., different-color balls, and square/circle/triangle-shaped goals in the game field.
}

\section*{Acknowledgement}
This work was supported by the National Science Foundation under Grants 2003211, 2128596, 2231209, 2146497, 2244219, 2315596, and 2413622.

\bibliographystyle{IEEEtran}


\section*{Biographies}
\scriptsize{
{\color{black}
\noindent {Yue Wang [SM]} is an Assistant Professor in the Department of Computer Science at Georgia State University. His interests include the areas of
machine learning, signal processing, and wireless communications, with research focuses on distributed optimization and 
learning, 
compressed sensing, Internet of Things, cognitive radios, and 5G/6G wireless techniques, and high-dimensional data analysis.
\vspace{0.06in}

\noindent {Zhi Tian [F]} is a Professor in the Department of Electrical and Computer Engineering at George Mason University. Her interests include 
signal processing, communications, detection and estimation with focuses on decentralized optimization and learning over networks, statistical inference from distributed data, and compressed sensing for random processes, cognitive radios, and millimeter-wave MIMO communications.
She is Editor-in-Chief for IEEE Transactions on Signal Processing. 
\vspace{0.06in}

\noindent {Xin Fan [M]} is an Assistant Professor with School of Information Science and Technology,
Beijing Forestry University. His interests include wireless communications, machine learning, security \& privacy, optimization, statistical signal processing, and blockchain.
\vspace{0.06in}

\noindent {Zhipeng Cai [F]} is a Professor in the Department of Computer Science at Georgia State University. His expertise lies in resource management and scheduling, high performance computing, cybersecurity, privacy, networking, and big data. He is Editor-in-Chief for Wireless Communications and Mobile Computing, and Associate Editor-in-Chief for Elsevier High-Confidence Computing Journal.
\vspace{0.06in}

\noindent {Cameron Nowzari [SM]} is an Associate Professor with the Department of Electrical and Computer Engineering at George Mason University. His 
interests include dynamical systems and control, distributed coordination algorithms, robotics, event- and self-triggered control, Markov processes, network science, spreading processes on networks, and Internet of Things.
\vspace{0.06in}

\noindent {Kai Zeng [M]} is a Professor with the Department of Electrical and Computer Engineering at George Mason University. 
His interests include cyber-physical system security and privacy, physical layer security, network forensics, and cognitive radio networks.}
}
\end{document}